\documentclass[twocolumn,english]{revtex4}
\usepackage[T1]{fontenc}
\usepackage[latin1]{inputenc}
\usepackage{graphicx}

\makeatletter

\usepackage{babel}
\makeatother
\begin{document}

\title{Domain wall formation and spin reorientation in finite-size magnetic
systems}

\author{R. M. Fernandes$^{\mathrm{1,2}}$, H. Westfahl Jr.$^{\mathrm{1}}$,
R. Magalh\~aes-Paniago$^{\mathrm{1,3}}$ and L. N. Coelho$^{\mathrm{1,3}}$}

\address{\emph{1) Laborat\'orio Nacional de Luz S\'{i}ncrotron, Caixa Postal 6192,
13084-971, Campinas, SP, Brazil}}

\address{\emph{2) Instituto de F\'{i}sica {}``Gleb Wataghin'', Universidade
Estadual de Campinas, 13083-970, Campinas, SP, Brazil}}

\address{\emph{3) Departamento de F\'{i}sica, Universidade Federal de Minas Gerais,
30123-970, Belo Horizonte, MG, Brazil}}

\date{10th October 2006}

\begin{abstract}
We investigate the formation of stable one-dimensional N\'eel walls
in a ferromagnetic slab with finite thickness and finite width. Taking
into account the dipolar, the exchange and the uniaxial anisotropic
crystalline field interactions, we derive an approximative analytical
self-consistent expression that gives the wall width in terms of ratios
between the three different energy scales of the problem. We also
show that, even when the crystalline anisotropy does not favour the
formation of domain walls, they can yet be formed due to the dipolar
interaction and the finiteness of the system. Moreover, using a Stoner-Wohlfarth
approach, we study the magnetization reorientation inside the domains
under the action of an external magnetic field and obtain the respective
hysteresis loops, showing that their shapes change from squared to
inclined as the width of the slab varies. Finally, we discuss possible
applications of this model to describe qualitatively some recent experimental
data on thin films of MnAs grown over GaAs substrates. 

\end{abstract}
\maketitle

\section{Introduction}

Magnetism in the micro and nanoscales is a source of promising technological
advances in a very broad range of interests, from spintronics and
quantum computation \cite{science2} to biophysics and pharmacology.
Nowadays, with the improvement of experimental methods for growth
and characterization of magnetic thin films, this particular class
of mesoscopic (quasi)two-dimensional systems has been largely investigated
\cite{surface1,surface2}. Particularly, one of the properties of
such films that has been calling more attention is their magnetic
domain structures and their dependence on temperature, film thickness
and applied magnetic field \cite{pappas,allen1,allen2,berger,mnas_PRL}.
Many of these systems present a competition between a short-range,
strong interaction (exchange) and a long-ranged, weak one (dipolar),
from which it is expected the emergence of spatially modulated configurations
\cite{science,garel}. However, rarely these theoretical models consider
possible size effects due to the finiteness of the film width; on
the contrary, usually they describe the films as infinite plates,
which is fair description of most experimental systems (see, for example,
\cite{teoria_filme1,teoria_filme2,teoria_filme3,teoria_filme4}).

One class of magnetic materials that cannot be described as an infinite
plate and that has been recently subjected to deep experimental analysis
are the thin films of MnAs grown over GaAs substrates (MnAs:GaAs)
\cite{mnas_PRL}. In bulk, MnAs exhibits a simultaneous abrupt first-order
magnetic/structural transition from a ferromagnetic, hexagonal phase
($\alpha$ phase) to a paramagnetic, orthorhombic one ($\beta$ phase)
\cite{bean}. However, MnAs:GaAs films do not show this abrupt transition;
conversely, a large region of coexistence between $\alpha$ and $\beta$
phases arises from $0\mathrm{^{^{\circ}}C}$ to $50\,^{\circ}\mathrm{C}$,
characterized by the formation of periodic $(\alpha+\beta)$ stripes
of constant width. The relative witdths of the $\alpha$ and $\beta$
phases varies with temperature while the total $(\alpha+\beta)$ width
remains constant \cite{engel,kaganer,ney,paniago,iikawa}. Hence,
to understand the domain structure inside each ferromagnetic stripe,
and how it varies with temperature, it is important to consider its
finite width, specially because in most experimental studies it is
of the same order of its thickness.

In this work, we apply the general method of energy minimization already
used in other contexts \cite{kittel,malek,aharoni} (such as infinite
plates, nanoparticles) to study the formation of stable unidimensional
domain walls in a slab with finite thickness and finite width. The
main purpose of such common procedure is to obtain an expression for
the total energy that includes the different interactions contributions.
In our case, there are three terms: the exchange term (which tends
to unfavour sharp walls); the uniaxial anisotropic crystalline term
(which can favour or not the formation of sharp walls, depending on
the easy axis of magnetization) and the dipolar term. This last one
is rather important in finite systems, as already pointed out by others
\cite{garel}; we show that, in the system considered here, it is
fundamental to form stable walls, specially when the crystalline anisotropy
does not favour them. From this energy expression, it is possible
to discuss the different solutions for the wall width depending on
the three scales of energy involved. 

We also generalize the energy expression to include domains whose
main magnetization axis is tilted by an angle $\phi$ with respect
to the normal direction to the film. With such expression, it is possible
to study how the magnetization is reorientated under the action of
an external magnetic field in the direction of the easy axis. This
is achieved by calculating theoretical hysteresis curves through a
method similar to the one proposed by Stoner and Wohlfarth \cite{stoner}.
This procedure does not take into account nucleation or pinning effects,
but only the rotation of the domains and may lead to values of magnetic
coercive fields that are not exactly the measured ones. However, as
it is a microscopic method, and not a phenomenological one, the main
properties predicted are expected to be followed by a variety of experimental
systems at least qualitatively. 

Here is an outline of the article: in Section 2, we propose a general
expression describing the magnetization corresponding to $N$ domains
whose walls width is $\sigma$ in a slab of thickness $D$ and width
$d$ and calculate the corresponding total energy, obtaining an approximative
analytical equation. Next, we minimize it with respect to $\sigma$
and discuss various possible solutions for the wall width depending
on the axis of magnetization (if it is the easy or the hard one) and
the relationship among the three distinct energy scales. In Section
3, we generalize the previous procedure to inclined domains and discuss
the different shapes of the hysteresis loops that describe spin reorientation
(i.e., if they are squared or inclined and the values of the coercive
fields). Section 4 is devoted to discuss a possible application of
the model developed to understand some properties of MnAs thin films
(particularly, recent experimental hysteresis curves). Section 5 contains
the conclusions and final remarks of the work.

\section{Energy minimization and wall width}

\begin{figure}
\begin{center}\includegraphics[%
  width=0.9\columnwidth,
  keepaspectratio]{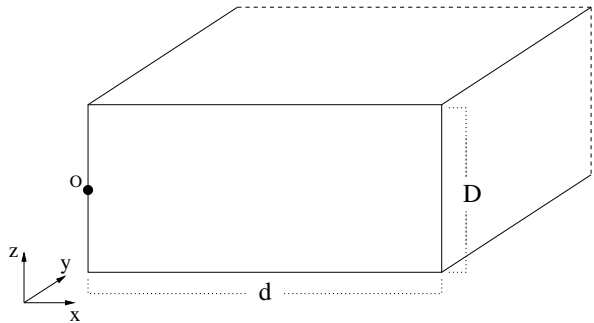}\end{center}

\caption{General picture of the coordinate system used to describe the ferromagnetic
slab.\label{fig:coord}}
\end{figure}

Through all this paper, we will use the coordinate system shown in
figure \ref{fig:coord}: the $z$ axis corresponds to the slab thickness
(which we shall call $D$), the $x$ axis, to the slab width (which
we shall call $d$) and the $y$ axis, to the slab length (considered
infinite for our purposes). The origin of the axes is located at the
middle point of one of the slab's faces, in a way that the $x$ axis
points in the direction of the other parallel face. We consider that
there is a strong crystalline field that does not allow the magnetization
to point in the $y$ direction ($xz$ spin model). This is the case
for many experimental systems and, specially, for the MnAs thin films.
But we will postpone the discussion about realistic applications of
the model until Section 4. 

Accordingly, the only kind of unidimensional domain walls that may
be formed in this system are N\'eel walls. Initially, we choose the
magnetization to lie along the $z$ direction, in such a way that
the walls are along the $x$ direction. Although we are considering
an uniaxial anisotropic system, we will not define yet which of the
axis is the easy one. To consider the role of the slab thickness on
the formation of these domains, we bound the magnetization to be nonzero
only inside it:

\begin{equation}
\vec{M}=\left[M_{x}(x)\hat{x}+M_{z}(x)\hat{z}\right]\theta(D/2+z)\theta(D/2-z)\label{magnet_total}\end{equation}
where $\theta(x)$ is the Heaviside step function. It can be shown
that this formulation would be equivalent to consider sharp domain
walls in the upper and lower borders of the slab such that their width
is much smaller than the width of the walls along the $x$ direction.
As we are concerned with the latter, there is no significant role
played by the former in what follows, and we can do $\sigma_{z}\rightarrow0$.

To investigate the formation of one-dimensional sharp domain
walls along the $x$ direction, we propose to study the configuration in which the $z$ component
of the magnetization is given by:

\begin{eqnarray}
M_{z}(x) & = & M_{0}\sum_{i=1}^{N}\frac{\left(-1\right)^{i-1}}{2}\left\{ \mathrm{erfc}\left[-\frac{x-d(i-1)/N}{\sqrt{2}\sigma}\right]\right.\nonumber \\
 &  & \left.-\mathrm{erfc}\left[-\frac{x-di/N}{\sqrt{2}\sigma}\right]\right\} \,,\label{Mz}\end{eqnarray}
where $M_{0}$ is the saturation magnetization, $\mathrm{erfc}(x)$
is the complementary error function, $N$ is the number of domains
inside the slab and $\sigma$ corresponds to the wall width along the $x$ axis, which
will be varied to minimize the total energy. This model resembles
the one used in reference \cite{aharoni}; the choice of using the
complementary error function is due to its analytical properties that
will allow us to obtain simple expressions for the wall width.

As we are considering a crystalline anisotropy such that $M_{y}=0$,
we can obtain $M_{x}$ using the fact that the module of the total
magnetization is constant and equal to $M_{0}$. Hence, if we meet
the condition:

\begin{equation}
\sigma\ll d\,,\label{condition}\end{equation}
then we can obtain a simple approximate expression for $M_{x}$ in
terms of Gaussian functions:

\begin{equation}
M_{x}(x)=M_{0}\sum_{i=1}^{N-1}e^{-\frac{(x-di/N)^{2}}{2\sigma^{2}}}\,.\label{Mx}\end{equation}

\begin{figure}
\includegraphics[%
  width=0.45\columnwidth]{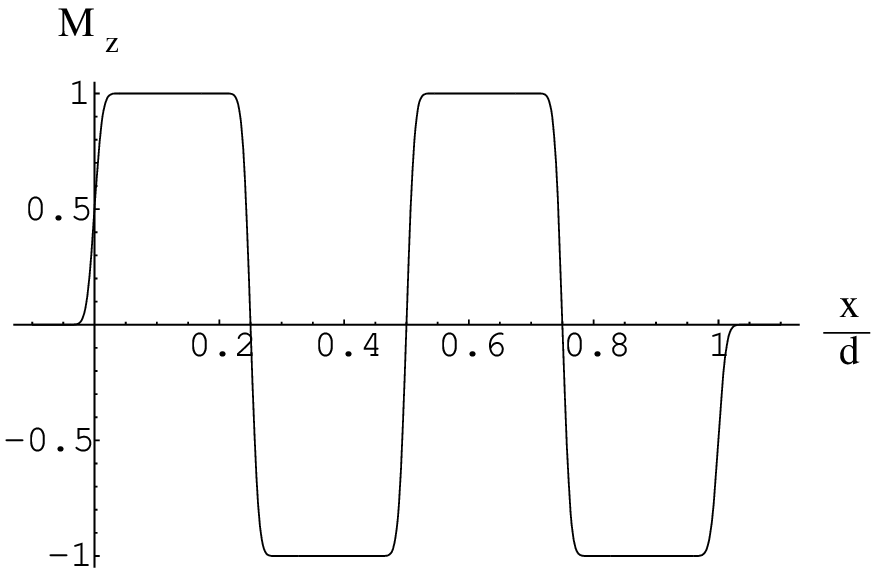}\hfill{}\includegraphics[%
  width=0.45\columnwidth]{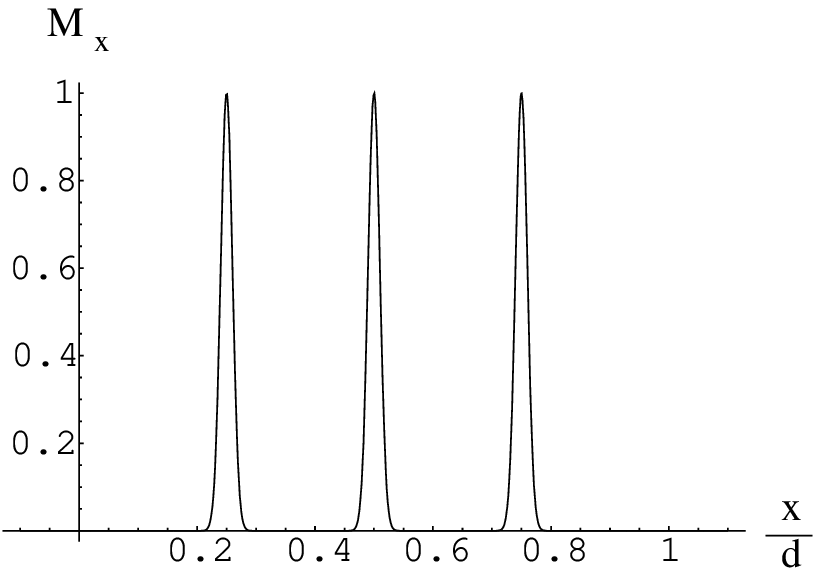}

~~~~~~~~~~~~~~~~~~~~~~~~~~(a)~~~~~~~~~~~~~~~~~~~~~~~~~~~~~~~~~~~~~~~~~~~~~~~~~~~~~~~~~~~~~~~~~~~~~~~~~~~~~~~~~(b)

\caption{\label{fig:configuration} Components (a) $M_{z}$ and (b) $M_{x}$
of the magnetization described by (\ref{Mz}) and (\ref{Mx}) for
a four domain configuration along the reduced slab's width $x/d$.}
\end{figure}

A typical domain configuration described by (\ref{Mz}) and (\ref{Mx})
is shown in figure \ref{fig:configuration}, where the $z$ and $x$
components of the magnetization are shown as a function of the slab's
width. 

As we discussed in the previous section, several works \cite{science,garel}
have shown that, in systems with competing interactions, a spatially
modulated configuration is expected. One may interpret these configurations
as {}``spread'' domain walls with sinusoidal domains; however, in
this article, we will investigate the formation of sharp walls, for
which condition (\ref{condition}) is expected to be satisfied. 

With the aid of expressions (\ref{Mz}) and (\ref{Mx}), we can calculate
the total energy of a specific domain configuration. However, it is
necessary to make an assumption about the microscopic nature of the
system if we want to include not only the {}``macroscopic'' terms
concerning the dipolar and the crystalline anisotropy interactions,
but also the {}``microscopic'' exchange energy. For long wavelengths,
one expects that the particular lattice structure (i.e., if it is
cubic, hexagonal, etc) will not substantially change the qualitative
physical properties derived for another kind of lattice. Hence, to
simplify the calculations, we follow \cite{garel} and choose the
system lattice to be cubic, with lattice parameter $a$ and gyromagnetic
factor $g$. Therefore, it is straightforward to relate the saturation
magnetization to the microscopic parameters:

\begin{equation}
M_{0}=\frac{g\mu_{B}}{a^{3}}\,,\label{saturation}\end{equation}
where $\mu_{B}$ is the Bohr's magneton. Now, it is possible to obtain
the three different energy terms. The exchange term is obtained from
the {}``classical'' Heisenberg nearest-neighbours Hamiltonian:

\begin{eqnarray*}
H_{exc} & = & -\frac{J}{2}\sum_{\left\langle ij\right\rangle }\vec{m}_{i}\vec{m}_{j}\\
 & = & \frac{J}{2a}\int d^{3}r\left[\left|\vec{\nabla}m_{x}(\vec{r})\right|^{2}+\left|\vec{\nabla}m_{z}(\vec{r})\right|^{2}\right]\,,\end{eqnarray*}
where we moved to the continuum limit of the lattice (see, for instance,
\cite{negele}) and defined $\vec{m}=\vec{M}/M_{0}$. Using (\ref{magnet_total}),
(\ref{Mz}) and (\ref{Mx}), we obtain that the exchange energy density
is given by:

\begin{equation}
\frac{E_{exc}}{V}=\frac{2J}{a^{3}}\left[\frac{2N+\pi(N-1)}{8\sqrt{\pi}(d/a)(\sigma/a)}\right]\,.\label{exchange}\end{equation}

To obtain the dipolar energy, as there are no free currents, we can
use equation (see, for instance, the Magnetostatics chapter of \cite{jackson}):

\begin{equation}
E_{dip}=\frac{1}{2}\int\phi(\vec{r})\rho(\vec{r})d^{3}r\,,\label{magnetostatic}\end{equation}
where $\rho=-\vec{\nabla}\cdot\vec{M}$ is the effective magnetic
charge density and $\phi(\vec{r})$ is the scalar magnetic potential,
which satisfies the Poisson equation:

\[
\nabla^{2}\phi=-4\pi\rho\,.\]

Taking the Fourier transforms of $\rho(\vec{r})$ and $\phi(\vec{r})$,
the Poisson equation can be easily solved and equation (\ref{magnetostatic})
can be written as:

\[
E_{dip}=2\pi\int\frac{\left|\rho(\vec{k})\right|^{2}}{k^{2}}d^{3}k\,.\]

Substituting expressions (\ref{magnet_total}), (\ref{Mz}) and (\ref{Mx}),
a straightforward calculation yields, for the magnetostatic energy
density:

\begin{eqnarray}
\frac{E_{dip}}{V} & = & \frac{2}{a^{3}}\left(\frac{g^{2}\mu_{B}^{2}}{a^{3}}\right)\left[\left(\frac{\sigma}{d}\right)^{2}\epsilon_{x}\left(\frac{\sigma}{d},\, N,\, p\right)\right.\nonumber \\
 &  & \left.+\epsilon_{z}\left(\frac{\sigma}{d},\, N,\, p\right)\right]\,,\label{dipolar1}\end{eqnarray}
where $p=d/D$ is the slab's aspect ratio and:

\begin{widetext}

\begin{eqnarray}
\epsilon_{x}\left(\frac{\sigma}{d},N,p\right) & = & 2\pi p\int_{0}^{\infty}du\frac{e^{-u^{2}\left(\frac{\sigma}{d}\right)^{2}}}{u}\left(\frac{u}{p}+e^{-u/p}-1\right)\left[1+\frac{\sin^{2}u/2}{\sin^{2}u/2N}-2\cos\left(\frac{u(N-1)}{2N}\right)\frac{\sin u/2}{\sin u/2N}\right]\nonumber \\
\epsilon_{z}\left(\frac{\sigma}{d},N,p\right) & = & 4p\int_{0}^{\infty}due^{-u^{2}\left(\frac{\sigma}{d}\right)^{2}}\frac{e^{-u/2p}}{u^{3}}\sinh\left(\frac{u}{2p}\right)\tan^{2}\left(\frac{u}{2N}\right)\left[1-(-1)^{N}\cos u\right]\,.\label{dipolar2}\end{eqnarray}

\end{widetext}

Finally, we can calculate the uniaxial crystalline anisotropic term
(see, for instance, the Magnetic Anisotropy chapter of \cite{aharoni}):

\[
E_{cryst}=-\Delta K\int d^{3}rm_{x}^{2}\,,\]
where $\Delta K$, the anisotropy constant, can be positive or negative,
depending if the $x$ axis is the easy ($\Delta K>0$) or the hard
one ($\Delta K<0$) . Evaluating this calculation, we obtain that
the density of anisotropic crystalline energy is:

\begin{equation}
\frac{E_{cryst}}{V}=-2\Delta K\frac{(N-1)\sqrt{\pi}}{2}\left(\frac{\sigma}{d}\right)\,.\label{crist}\end{equation}

Hence, the total energy is given by:

\begin{equation}
E_{tot}=E_{exc}+E_{dip}+E_{cryst}\label{energy_total}\end{equation}

As we are assuming condition (\ref{condition}) to be satisfied, we
can make a further approximation to obtain a simpler expression for
the total energy. Using such condition, we can approximate the exponential
$e^{-u^{2}(\sigma/d)^{2}}$ in the integrals (\ref{dipolar2}) to
$1$, as long as we take an appropriate upper limit to them:

\begin{equation}
\epsilon_{x}\left(\frac{\sigma}{d},N,p\right)\approx\epsilon_{x}\left(0,N,p\right)\qquad\epsilon_{z}\left(\frac{\sigma}{d},N,p\right)\approx\epsilon_{z}\left(0,N,p\right)\label{aprox}\end{equation}

This procedure implies in errors of the order of $10\%$ to $20\%$,
if compared to numerical calculations. The choice of the upper limit
of the integrals has a small impact on the final result since we are
going to apply a self-consistent method in the end of the calculation.
Therefore, the qualitative physical properties are still valid in
this approximation, and we can minimize (\ref{energy_total}) as:\begin{eqnarray*}
\frac{\partial E_{tot}}{\partial\sigma} & = & 0\\
\frac{\partial^{2}E_{tot}}{\partial\sigma^{2}} & > & 0\,.\end{eqnarray*}

To solve these equations, it is convenient to use the following auxiliary
variables:

\begin{widetext}

\begin{eqnarray}
\left(\frac{\sigma'}{a}\right) & = & \left(\frac{\sigma}{a}\right)\left(\frac{J}{g^{2}\mu_{B}^{2}/a^{3}}\right)^{-1/3}\left(\frac{2N+\pi(N-1)}{64\sqrt{\pi}\left(a/d\right)\epsilon_{x}\left(0,N,p\right)}\right)^{-1/3}\nonumber \\
\lambda & = & \left[\frac{\left(\Delta Ka^{3}\right)^{3}}{J\left(g^{2}\mu_{B}^{2}/a^{3}\right)^{2}}\right]^{1/3}\left[\frac{2(N-1)^{3}\pi^{2}}{27\left(a\epsilon_{x}\left(0,N,p\right)/d\right)^{2}\left(2N+\pi(N-1)\right)}\right]^{1/3}\,.\label{aux_var}\end{eqnarray}

\end{widetext}

Then, the equation for the domain wall width can be written as:

\begin{equation}
\left(\frac{\sigma'}{a}\right)^{3}-3\lambda\left(\frac{\sigma'}{a}\right)^{2}-4=0\,.\label{aux_equation}\end{equation}

It is easy to see that the auxiliary variables (\ref{aux_var}) are
just relations between the three different energy scales involved
in the system: in $\sigma'$, there is the ratio between the typical
value of the exchange energy and the typical value of the dipolar
energy. The other term is just a numerical one and depends only on
the ratio $a/d$, the number of domains $N$ and the aspect ratio
$p$. For thickness of the order of hundreds of lattice parameters,
this numerical factor is usually of the order of $10^{-1}$, what
implies that the relationship between the typical exchange and dipolar
energies will determine the order of the wall width. From this, it
is clear that when $J\gg g^{2}\mu_{B}^{2}/a^{3}$ no sharp walls would
be formed, as it would be expected.

The parameter $\lambda$ is a relationship between the three types
of energy and can be positive or negative, depending if the $x$ axis
is the easy or the hard one, respectively. The numerical factor again
depends only on $a/d$, $N$ and $p$, and for thickness of the order
of hundreds of lattice parameters, it is usually of the order of $10^{-1}$. 

Let us study the solutions of (\ref{aux_equation}); independently
of the sign of $\lambda$, equation (\ref{aux_equation}) always has
only one positive solution. If $\lambda\geq0$, this solution is:

\begin{eqnarray}
\left(\frac{\sigma'}{a}\right) & = & \lambda+\frac{\lambda^{2}}{\left(1+\sqrt{1+\lambda^{3}}\right)^{2/3}}\label{sol_positive}\\
 &  & +\left(1+\sqrt{1+\lambda^{3}}\right)^{2/3}\,,\nonumber \end{eqnarray}
while, for $-1\leq\lambda<0$, we have:

\begin{eqnarray}
\left(\frac{\sigma'}{a}\right) & = & \lambda+\frac{\lambda^{2}}{\left(1-\sqrt{1+\lambda^{3}}\right)^{2/3}}\label{sol_negative>-1}\\
 &  & +\left(1-\sqrt{1+\lambda^{3}}\right)^{2/3}\,,\nonumber \end{eqnarray}
and in the case where $\lambda<-1$:

\begin{eqnarray}
\left(\frac{\sigma'}{a}\right) & = & \lambda-2\lambda\cos\left[\frac{\arg\left(2+\lambda^{3}-2\sqrt{\lambda^{3}+1}\right)}{3}\right]\label{sol_negative<-1}\end{eqnarray}

\begin{figure}
\begin{center}\includegraphics[%
  width=0.9\columnwidth]{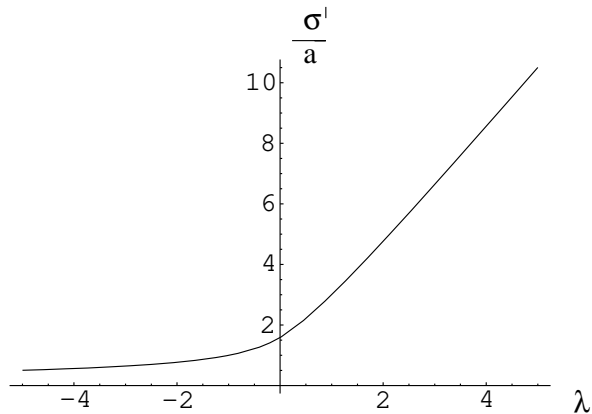}\end{center}

\caption{Plot of the general solution $\sigma'/a$ for the cubic equation
(\ref{aux_equation}) as a function of the parameter $\lambda$.\label{fig:sol_cub}}
\end{figure}

In figure \ref{fig:sol_cub}, we show the graphics of the positive
solution as a function of the parameter $\lambda$. As expected, when
$\lambda$ is negative (magnetization lying on the easy axis), the
walls width is smaller than when $\lambda$ is positive (magnetization
lying on the hard axis). We note that even when $\lambda>0$, i.e.,
the crystalline anisotropy does not favour the formation of walls,
it is possible for the system to be divided in stable domains, as
long as the dipolar interaction is large enough compared to both the
other two energy scales (as they appear in (\ref{aux_var})). 

An interesting situation that deserves a deeper analysis is when $0<\lambda\ll1$.
Such case would occur, for example, if the energy scales satisfied
$J\sim g^{2}\mu_{B}^{2}/a^{3}$ and $g^{2}\mu_{B}^{2}/a^{3}\gg\Delta Ka^{3}$.
Then, we can expand (\ref{sol_positive}) and obtain:

\[
\left(\frac{\sigma'}{a}\right)=2^{2/3}+\lambda+\mathcal{O}\left(\lambda^{2}\right)\,,\]
from which we get, to first order in $\lambda$:

\begin{eqnarray}
\left(\frac{\sigma}{a}\right) & = & \left(\frac{a^{3}J}{g^{2}\mu_{B}^{2}}\right)^{1/3}\left(\frac{2N+\pi(N-1)}{16\sqrt{\pi}\left(\frac{a}{d}\right)\epsilon_{x}\left(0,N,p\right)}\right)^{1/3}\nonumber \\
 & + & \left(\frac{\Delta K\, a^{6}}{g^{2}\mu_{B}^{2}}\right)\frac{(N-1)\sqrt{\pi}}{\frac{12a}{d}\epsilon_{x}\left(0,N,p\right)}\,.\label{sol_lambda<<1}\end{eqnarray}

To avoid numerical problems and reduce the errors concerning the approximation
(\ref{aprox}), it is useful to transform the general solution for
the walls width in a self-consistent equation:

\begin{eqnarray}
\left(\frac{\sigma}{a}\right) & = & \left(\frac{\sigma'}{a}\right)\left(\frac{a^{3}J}{g^{2}\mu_{B}^{2}}\right)^{1/3}\times\label{self_consistent}\\
 &  & \left(\frac{2N+\pi(N-1)}{64\sqrt{\pi}\frac{a}{d}\,\epsilon_{x}\left(\frac{\sigma}{d},N,p\right)}\right)^{1/3}\,,\nonumber \end{eqnarray}
in which there is an implicit dependence of $\sigma'/a$ upon $\sigma$.
Although this procedure does not give exactly the same result as a
numerical method, it prevents one from making further errors due to
the integral that appears in $\epsilon_{x}\left(\frac{\sigma}{d},N,p\right)$.

Finally, it is interesting to study this model in the limit of $d,D\rightarrow\infty$,
where one expects to recover the result known as Landau-Lifshitz wall
(see, for instance \cite{aharoni}). From equation (\ref{energy_total}),
it is clear that the terms referring to the exchange and to the crystalline
anisotropy are proportional to $d^{-1}$. The dipolar term has two
parts: $\epsilon_{z}$, which is due to the surface charge on the
upper and lower faces of the slab, goes to zero as $D\rightarrow\infty$;
$\epsilon_{x}$ is just a finite number for a constant aspect ratio
$p$. Then, making $p$ constant as $d,D\rightarrow\infty$, it is
clear that this term is proportional to $d^{-2}$ and is negligible
if compared to the others. Hence, as pointed in \cite{aharoni}, the
dipolar energy vanishes in the limit of an {}``infinite'' crystal
and domain walls will be formed only if $\Delta K<0$. Minimizing
the energy with respect to $\sigma$ and putting $N=2$, we find the
result:

\[
\left(\frac{\sigma}{a}\right)=1.06\sqrt{\frac{J}{2\left|\Delta K\right|a^{3}}}\,,\]
which is just $6\%$ greater than the value of the Landau-Lifshitz
wall.

\section{Theoretical hysteresis loops}

We now want to study how this system responds to an external magnetic
field applied along the $x$ axis. It is expected that, for strong
enough fields, the magnetization will lie along the $x$ axis and
will follow the field direction. Our main objective is to determine
the qualitative features of the hysteresis curves that would be observed
for different slab widths. Therefore, we must generalize our previous
model to include rotations of the domains magnetization. Introducing
the angle $\phi$ between the magnetization and the $z$ axis, we
can write its components approximately as:

\begin{widetext}

\begin{eqnarray}
M_{z}(x) & = & M_{0}\cos\phi\sum_{i=1}^{N}\frac{\left(-1\right)^{i-1}}{2}\left\{ \mathrm{erfc}\left[-\frac{x-d(i-1)/N}{\sqrt{2}\sigma}\right]-\mathrm{erfc}\left[-\frac{x-di/N}{\sqrt{2}\sigma}\right]\right\} \nonumber \\
M_{x}(x) & = & M_{0}\left(1-\sin\phi\right)\sum_{i=1}^{N-1}e^{-\frac{(x-di/N)^{2}}{2\sigma^{2}}}+M_{0}\frac{\sin\phi}{2}\left\{ \mathrm{erfc}\left[-\frac{x}{\sqrt{2}\sigma}\right]-\mathrm{erfc}\left[-\frac{x-d}{\sqrt{2}\sigma}\right]\right\} \,,\label{magnet_phi}\end{eqnarray}

\end{widetext} assuming that the condition (\ref{condition}) is
satisfied. In what follows, we will consider that the width of the
walls is constant, to simplify the calculations. Using the formalism
of last section, it is straightforward to conclude that an external
magnetic field along the $x$ axis usually will not substantially
change the value of the minimum wall width.

It is important to notice that this expression for $M_{x}$, (\ref{magnet_phi}),
as well as the previous one, (\ref{Mx}), assumes that the domains
are along the positive $x$ direction. Although this feature does
not influence any of the results obtained in the last section, it
must be taken into account in this one, since we are dealing with
spin rotation. If we were to make a complete description of this phenomenon,
it would be necessary to include domains in the $y$ direction, otherwise
the magnetization inside the walls could never rotate properly. Hence,
bidimensional domains would have to be considered, but this is beyond
the scope of this article. Therefore, as the domain walls are very
small, we will not treat their rotation, but only the domains rotation.
This procedure will then be enough to give us the main qualitative
characteristics of the hysteresis loops.

Repeating the procedure of the last section, it is straightforward
to calculate the total energy density referring to the configuration
(\ref{magnet_phi}). We substitute it in the expressions for the
exchange, dipolar and crystalline anisotropic energies, obtaining the total
energy according to (\ref{energy_total}). Leaving only the terms proportional to $\phi$,
we obtain:

\begin{eqnarray}
\frac{E}{2M_{0}^{2}} & = & A\left(N,d,D\right)\sin^{2}\phi\label{energ_phi>0}\\
 &  & -\left[B\left(N,d,D\right)+\frac{H}{M_{0}}C\left(N,\frac{\sigma}{d}\right)\right]\sin\phi\,,\nonumber \end{eqnarray}

where $H$ is the external magnetic field along the $x$ direction and:

\begin{widetext}

\begin{eqnarray}
A\left(N,d,D\right) & = & 2\pi p\int_{0}^{\infty}\frac{e^{-u^{2}\left(\frac{\sigma}{d}\right)^{2}}}{u}\left(\frac{u}{p}+e^{-\frac{u}{p}}-1\right)\left\{ \left(\frac{\sigma}{d}\right)^{2}\left[1+\frac{\sin^{2}u/2}{\sin^{2}u/2N}-2\cos\left(\frac{u(N-1)}{2N}\right)\frac{\sin\frac{u}{2}}{\sin\frac{u}{2N}}\right]\right.\nonumber \\
 &  & \left.+\frac{2}{\pi}\frac{\sin^{2}u/2}{u^{2}}-\left(\frac{\sigma}{d}\right)\sqrt{\frac{2}{\pi}}\frac{\sin u/2}{u}\left[\cot\left(\frac{u}{2N}\right)\tan\left(\frac{u}{2}\right)-1\right]du\right\} \nonumber \\
 &  & -4p\int_{0}^{\infty}e^{-u^{2}\left(\frac{\sigma}{d}\right)^{2}}\frac{e^{-u/2p}}{u^{3}}\sinh\left(\frac{u}{2p}\right)\tan^{2}\left(\frac{u}{2N}\right)\left[1-(-1)^{N}\cos u\right]du\nonumber \\
 &  & +\left(\frac{J}{g^{2}\mu_{B}^{2}/a^{3}}\right)\left[\frac{-2N+\pi(N-1)+2}{8\sqrt{\pi}(d/a)(\sigma/a)}\right]-\frac{1}{2}\left(\frac{\Delta Ka^{3}}{g^{2}\mu_{B}^{2}/a^{3}}\right)\nonumber \\
B\left(N,d,D\right) & = & 2\pi p\int_{0}^{\infty}\frac{e^{-u^{2}\left(\frac{\sigma}{d}\right)^{2}}}{u}\left(\frac{u}{p}+e^{-\frac{u}{p}}-1\right)\left\{ \left(\frac{\sigma}{d}\right)^{2}\left[1+\frac{\sin^{2}u/2}{\sin^{2}u/2N}-2\cos\left(\frac{u(N-1)}{2N}\right)\frac{\sin\frac{u}{2}}{\sin\frac{u}{2N}}\right]\right.\nonumber \\
 &  & \left.-\left(\frac{\sigma}{d}\right)\frac{1}{2\pi}\frac{\sin u/2}{u}\left[\cot\left(\frac{u}{2N}\right)\tan\left(\frac{u}{2}\right)-1\right]du\right\} \nonumber \\
 &  & +2\left(\frac{J}{g^{2}\mu_{B}^{2}/a^{3}}\right)\left[\frac{\pi(N-1)}{8\sqrt{\pi}(d/a)(\sigma/a)}\right]\nonumber \\
C\left(N,\frac{\sigma}{d}\right) & = & \frac{1}{2}\left[1-\left(\frac{\sigma}{d}\right)\sqrt{2\pi}(N-1)\right]\,.\label{parametros_stoner}\end{eqnarray}

\end{widetext}

If the condition (\ref{condition}) is satisfied, one can usually
make the approximation $C\approx1/2$, as long as there are few domains
inside the slab. 

As we did not consider the terms independent of $\phi$ in the total
energy expression (\ref{energ_phi>0}), we note that the energy of the configuration studied
in the previous section, corresponding to $\phi=0$, would be $E=0$. Indeed, the terms that appear in (\ref{energ_phi>0}) are a
combination of the exchange, dipolar and crystalline anisotropic energies referring to the components
of the magnetization (\ref{magnet_phi}) that are not parallel to the $z$ direction. It is clear that the total
energy will be non-zero only for configurations in which the the magnetization is tilted by some angle with respect to the $z$ axis.

Now, it is possible to apply a procedure similar to the one developed
by Stoner and Wohlfarth \cite{stoner} to analyze the spin reorientation
in the presence of a magnetic field. Minimizing equation (\ref{parametros_stoner})
with respect to $\phi$ leads us to 

\begin{eqnarray}
\frac{\partial E}{\partial\phi} & = & 2A\sin\phi\cos\phi-\left(B+\frac{H}{M_{0}}C\right)\cos\phi\label{first_deriv}\\
 & = & 0\,,\nonumber \end{eqnarray}
while the second derivative is given by

\begin{equation}
\frac{\partial^{2}E}{\partial\phi^{2}}=2A\cos2\phi+\left(B+\frac{H}{M_{0}}C\right)\sin\phi\,.\label{second_deriv}\end{equation}

Equation (\ref{first_deriv}) has two possible solutions: the first
one refers to the two possible orientations for which the magnetization
lies on the $x$ axis:

\begin{equation}
\cos\phi_{1}=0\Rightarrow\phi_{1}=\pm\frac{\pi}{2}\,,\label{sol1_phi_+-}\end{equation}
and is a minimum as long as:

\begin{equation}
-2A+\left(B+\frac{H}{M_{0}}C\right)\geq0\,,\label{estab_+pi/2}\end{equation}
if $\phi_{1}=\pi/2$~, or:

\begin{equation}
-2A-\left(B+\frac{H}{M_{0}}C\right)\geq0\,,\label{estab_-pi/2}\end{equation}
if $\phi_{1}=-\pi/2$. The second solution is:

\begin{equation}
\phi_{2}=\arcsin\left(\frac{B+\frac{H}{M_{0}}C}{2A}\right)\,,\label{sol2_phi}\end{equation}
and it is a minimum for

\begin{equation}
2A\geq\frac{\left(B+\frac{H}{M_{0}}C\right)^{2}}{2A}\,.\label{estab_sol2}\end{equation}

Therefore, we can build a procedure to trace the upper curve of the
hysteresis loop. Applying a strong enough positive magnetic field,
the magnetization will lie along the positive $x$ direction, since
$\phi_{1}=\pi/2$ will be the global minimum. As $H$ diminishes,
and eventually becomes negative, condition (\ref{estab_+pi/2}), at
some moment, will no more be satisfied and two options will raise.
If $A\leq0$, while condition (\ref{estab_sol2}) will never be satisfied,
condition (\ref{estab_-pi/2}) will be, and the system will jump to
the configuration in which the magnetization lies along the negative
$x$ direction ($\phi=-\pi/2$). Hence, the upper curve of the hysteresis
loop has the shape of an abrupt step when $A\leq0$. The remanent
magnetization along the $x$ axis is simply

\begin{equation}
M_{r}=M_{0}\,,\label{remanent_squared}\end{equation}
and the coercive field is given by:

\begin{equation}
H_{c}=-M_{0}\left(\frac{B-2A}{C}\right)\,.\label{coercive_squared}\end{equation}

However, if $A>0$, other stable states will raise before $\phi=-\pi/2$
is reached, since condition (\ref{estab_sol2}) is satisfied. Then,
the system jumps continuously from minimum to minimum, as none of
them can be a metastable state. This behaviour continues until $H$
is strong enough to not satisfy condition (\ref{estab_sol2}) anymore.
Hence, condition (\ref{estab_-pi/2}) is satisfied and the magnetization
lies along the negative $x$ direction. In this case, in which $A>0$,
the upper curve has the shape of an inclined step. The remanent magnetization
along the $x$ axis is given by:

\begin{equation}
M_{r}=\left\{ \begin{array}{ll}
M_{0} & ,\;\mathrm{if}\; B>2A\\
M_{0}\left(\frac{B}{2A}\right) & ,\;\mathrm{otherwise}\end{array}\right.\label{remanent_inclined}\end{equation}
and the coercive field by

\begin{equation}
H_{c}=-M_{0}\left(\frac{B}{2C}\right)\,.\label{coercive_inclined}\end{equation}

To obtain the lower curve of the hysteresis loop, we have to perform
a slight modification. As previously explained, the configuration
considered assumes that the domain walls are along the positive $x$
direction. However, after the magnetization is reoriented, we must
consider a new configuration in which the domain walls are along the
negative $x$ direction, so we avoid additional errors for not considering
domains along the $y$ direction. Hence, we take:

\begin{widetext}

\begin{equation}
M_{x}(x)=M_{0}\left(-1-\sin\phi\right)\sum_{i=1}^{N-1}e^{-\frac{(x-di/N)^{2}}{2\sigma^{2}}}+M_{0}\frac{\sin\phi}{2}\left\{ \mathrm{erfc}\left[-\frac{x}{\sqrt{2}\sigma}\right]-\mathrm{erfc}\left[-\frac{x-d}{\sqrt{2}\sigma}\right]\right\} \,.\label{mx_hyster_2}\end{equation}

\end{widetext}

Equation (\ref{energ_phi>0}) is then replaced by

\begin{widetext}\begin{equation}
\frac{E}{2M_{0}^{2}}=A\left(N,d,D\right)\sin^{2}\phi+\left[B\left(N,d,D\right)-\frac{H}{M_{0}}C\left(N,\frac{\sigma}{d}\right)\right]\sin\phi\,.\label{energ_phi<0}\end{equation}

\end{widetext}

Following the same procedure as before, we obtain that the lower curve
has the same shape of the upper one, and the remanent magnetization
along the $x$ axis and the coercive field are symmetrical to the
ones presented previously, (\ref{remanent_squared})-(\ref{coercive_inclined}). 

Therefore, we conclude that, if $A\leq0$, the hysteresis loop is
squared, the remanent magnetization is $M_{0}$ and the coercive field
is given by (\ref{coercive_squared}). If $A>0$, instead, the hysteresis
loop is inclined, the remanent magnetization is given by (\ref{remanent_inclined})
and the coercive field by (\ref{coercive_inclined}).

\section{Application to $\mathrm{MnAs:GaAs}$}

Several works were done regarding magnetic properties of MnAs thin
films grown over GaAs substrates. Here we will discuss possible applications
of our model to the experimental hysteresis loops recently obtained
\cite{histerese_grao,histerese_ploog,histerese_nosso}. To make correspondence
between our model and the real system, it is useful to identify the
following crystallographic directions of MnAs to the axis considered
in figure 1: $[\bar{1}\bar{1}20]=x$, $[0001]=y$ and $[\bar{1}100]=z$.

First of all, let us outline the parameters of the MnAs thin films
that we will use in our calculations. As measured in \cite{lindner},
the anisotropy constants associated to the $z$ and to the $x$ axis
are, respectively, $K_{z}=7\cdot10^{6}\,\mathrm{erg/cm^{3}}$ and
$K_{x}=7.4\cdot10^{6}\,\mathrm{erg/cm^{3}}$. Hence, the $x$ axis
is the easy one, and we can use $\Delta K=0.4\cdot10^{6}\,\mathrm{erg/cm^{3}}$.
From the crystalline structure of MnAs \cite{mnas_PRL},which is in
fact hexagonal, we estimate the value of the equivalent cubic lattice
parameter to be $a=4\,\mathrm{A}$. The gyromagnetic factor was estimated
as $g=4.5$, from which we obtain a magnetization of $M_{0}=0.65\cdot10^{6}\,\mathrm{A/m}$,
that is very close to the experimental value measured of $M_{0}=0.67\cdot10^{6}\,\mathrm{A/m}$
\cite{mnas_M0}. As we are interested in orders of magnitude, we estimate
$J=4.5\,\mathrm{meV}$ from the Curie temperature of MnAs. Finally,
as these films have constant thickness, we take the fixed value $D=130\,\mathrm{nm}$,
which is the thickness of the sample used in \cite{histerese_nosso}
and of the same order of magnitude of usual samples.

Magnetic Force Microscopy (MFM) images of MnAs:GaAs films have suggested
that \cite{plake,histerese_nosso}, as the temperature rises in the
coexistence region (implying in smaller widths of the ferromagnetic
stripe), the system undergoes a transition from a configuration in
which the magnetization lies along the easy $x$ axis to another one
in which it lies along the growth axis (the hard $z$ axis) and is
divided in three domains. Let us verify the predictions of our model
for these situation: substituting the experimental parameters of MnAs
in the self-consistent equation (\ref{self_consistent}) for $N=3$
domains, we obtain that the domain walls are stable and that their
width is given by $\sigma=3.8a$. The solution to the cubic equation
(\ref{aux_equation}) used was (\ref{sol_lambda<<1}), since the experimental
parameters of MnAs imply $\lambda\ll1$. A numerical calculation in
fact gives a width of about $5a$, what confirms our expectations
that the approximative self-consistent method gives errors less than
$20\%$.

\begin{figure}
\begin{center}\includegraphics[%
  width=0.9\columnwidth]{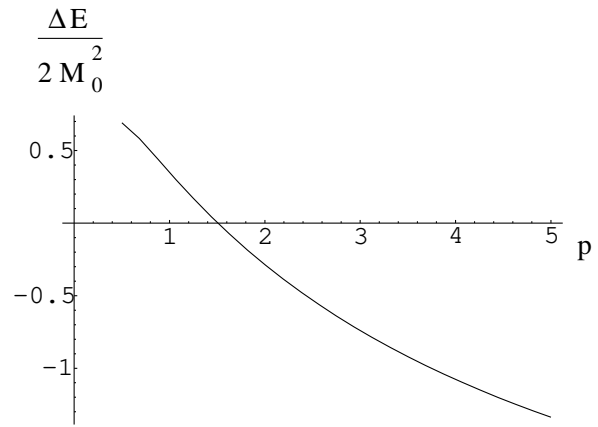}\end{center}

\caption{Plot of the energy difference between the configuration
in which the magnetization lies along the $x$ axis and the one in
which there are three domains lying along the $z$ axis as a function
of the aspect ratio $p$ of the ferromagnetic stripe. \label{fig:energy}}
\end{figure}

Using equation (\ref{energ_phi>0}), it is possible to obtain the
energy difference between the two configurations seen in the MFM images.
In figure \ref{fig:energy}, we show the plot of this difference as
a function of the aspect ratio $p$ and notice that there is a transition
at $p_{c}\approx1.5$. The measurements realized by Coelho \emph{et.
al.} \cite{histerese_nosso} suggest that this transition takes place
around $p_{c}\approx2.9$. However, as mentioned previously, we do
not take into account the domains along the $y$ direction or the
inter-stripe interaction, a feature always observed in the MFM images.
This seems to be particular important not only to provide more precise
values for $p_{c}$ but also to ensure that the most stable configuration
has $3$ domains. Without taking into account such features, our model
would predict also transitions to $N>3$ domain configurations before
the one at $N=3$, what is not observed in the MFM images. Hence,
we can only predict a qualitative behaviour of the real system.

\begin{figure}
\begin{center}\includegraphics[%
  width=0.9\columnwidth]{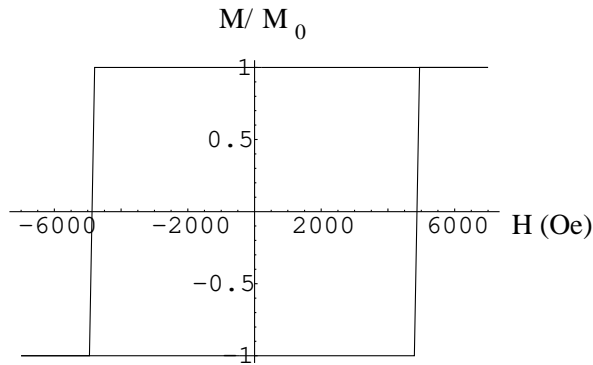}\end{center}

\caption{\label{p19}Plot of the hysteresis loop predicted by the model for
$p=1.9$. The vertical axis represents the relative magnetization
$M/M_{0}$ along the $x$ axis and the horizontal axis represents
the external magnetic field $H$ applied in the $x$ direction in
$\mathrm{Oe}$.\label{fig:hyst1}}
\end{figure}

\begin{figure}
\begin{center}\includegraphics[%
  width=0.9\columnwidth]{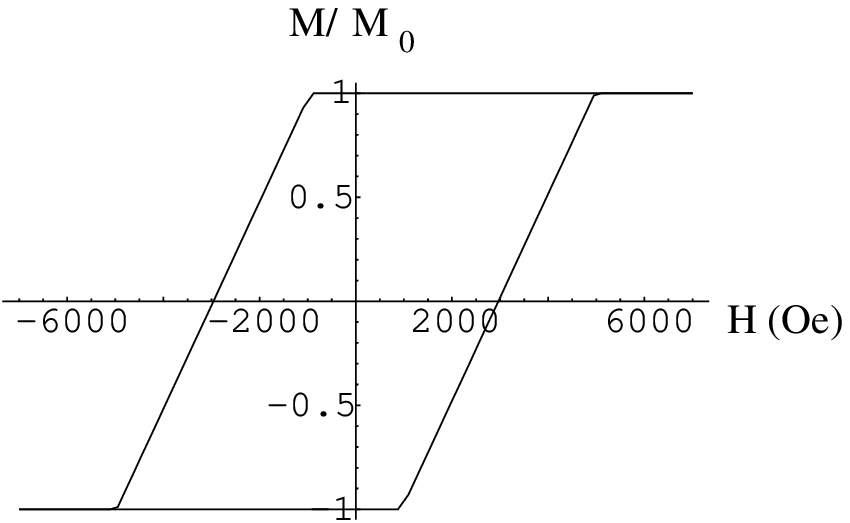}\end{center}

\caption{\label{p17}Plot of the hysteresis loop predicted by the model for
$p=1.7$. The vertical axis represents the relative magnetization
$M/M_{0}$ along the $x$ axis and the horizontal axis represents
the external magnetic field $H$ applied in the $x$ direction in
$\mathrm{Oe}$.\label{fig:hyst2}}
\end{figure}

\begin{figure}
\begin{center}\includegraphics[%
  width=0.9\columnwidth]{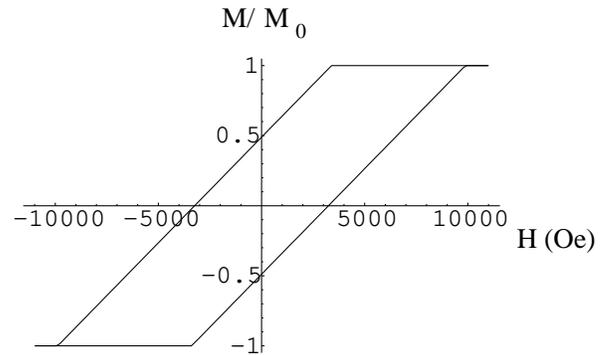}\end{center}

\caption{\label{p15}Plot of the hysteresis loop predicted by the model for
$p=1.5$. The vertical axis represents the relative magnetization
$M/M_{0}$ along the $x$ axis and the horizontal axis represents
the external magnetic field $H$ applied in the $x$ direction in
$\mathrm{Oe}$.\label{fig:hyst3}}
\end{figure}

It is interesting to plot the theoretical hysteresis loops predicted
by our model for a $N=3$ domain configuration with the MnAs experimental
parameters. Applying the procedure developed in the previous section,
we see that $B>0$ for any value of the aspect ratio $p$ and $A$
is negative until $p\approx1.8$ , where it becomes positive. Hence,
it is expected that the hysteresis shape is squared until $p\approx1.8$,
where it becomes inclined. For $1.65<p<1.8$, the remanent magnetization
is expected to be yet $M_{0}$, and only for $p<1.65$ it will start
to decrease. Figures \ref{fig:hyst1}, \ref{fig:hyst2} and \ref{fig:hyst3}
show the three different possible shapes of the hysteresis loops,
according to the procedure of the last section. 

\begin{figure}
\includegraphics[%
  width=0.45\columnwidth]{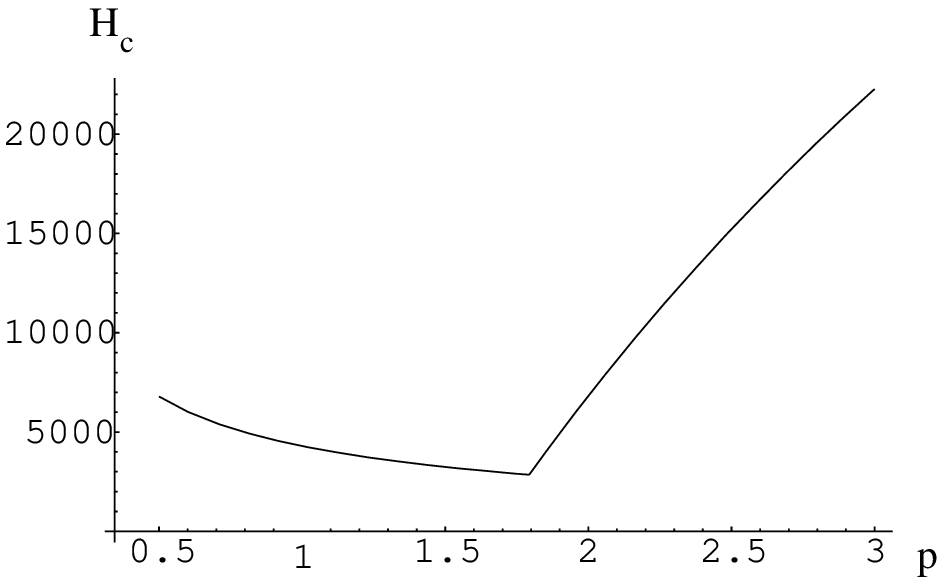}\hfill{}\includegraphics[%
  width=0.45\columnwidth]{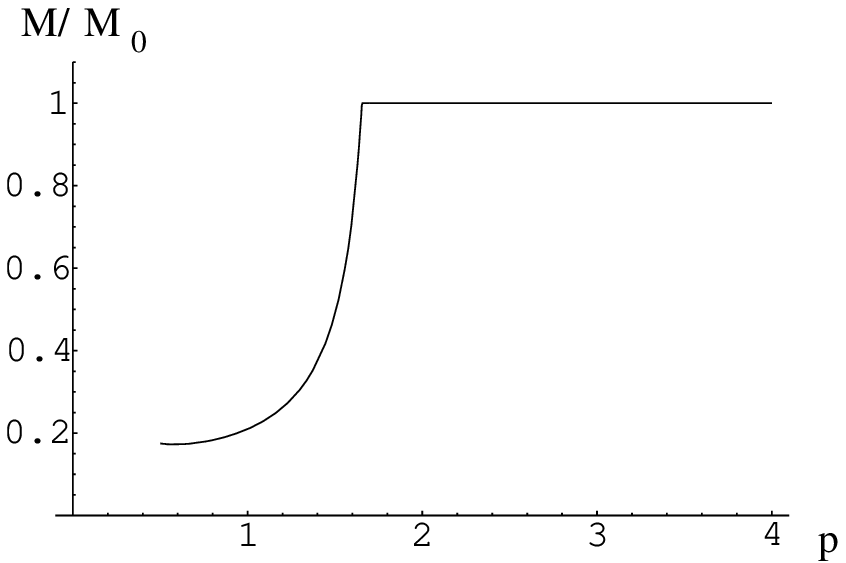}

~~~~~~~~~~~~~~~~~~~~~~~~~~(a)~~~~~~~~~~~~~~~~~~~~~~~~~~~~~~~~~~~~~~~~~~~~~~~~~~~~~~~~~~~~~~~~~~~~~~~~~~~~~~~~~(b)

\caption{Plot of the (a) modulus $H_{c}$ of the coercive field (in $\mathrm{Oe}$)
and (b) of the relative remanent magnetization $M/M_{0}$ along the
$x$ axis as a function of the aspect ratio $p$ of the ferromagnetic
stripe.\label{fig:coerc}}
\end{figure}

It is worth to notice that hysteresis loops of shapes similar to figures
\ref{p15} and \ref{p17} (but more rounded) were observed in \cite{histerese_ploog}
and \cite{histerese_nosso}, respectively, while shapes like the one
in figure 4 were seen in both of them. Coelho \emph{et. al.} showed
that the change in the shape of the hysteresis loops occurs for $p\approx2.9$,
which is not too far from our prediction ($p\approx1.8$). However,
our coercive fields are one order of magnitude greater than the values
obtained by both \cite{histerese_nosso} and \cite{histerese_ploog}.
It is interesting to notice that simulations performed by Engel-Herbert
\emph{et. al.} \cite{histerese_ploog} considering other configurations
and other methods led to the same order of magnitude for the coercive
fields than our model. The graphics of the coercive field and the
remanent magnetization predicted by our model are shown in figure
\ref{fig:coerc}, where it is easy to verify the paths for the transitions
among the three different types of hysteresis loops pointed before. 

Finally, we point out that, in \cite{histerese_grao}, Takagaki \emph{et.
al.} found a hysteresis loop similar to the squared one (figure \ref{p19})
for MnAs thin films and another similar to the inclined one (figure
\ref{p17}) (but more rounded, again) for disks of MnAs fabricated
from thin films. It is evident that our model cannot be applied to
disks (in such case, there is the possibility of more complex configurations,
like vortices, for example), but it may give a hint about their physical
behaviour.

\section{Conclusions}

In this article, we have shown, using an approximative analytical
self-consistent equation, that multiple-domains configurations with
sharp walls in a ferromagnetic slab can be stable. We arrived at ratios
among the three typical magnetic energy scales of the system (exchange,
dipolar and crystalline anisotropic) that can determine the stability
of these configurations and also give the order of magnitude for the
walls width. Although the approximations done to achieve these equations
introduces imprecision on the predictions of the model, they do not
significantly change the orders of magnitude involved. Moreover, this
model shows that, even when the crystalline anisotropy prefers spread
walls, the dipolar interaction can compensate it to form sharp ones.
In what concerns thin films of MnAs grown over GaAs substrates, which
could be an observable realization of our model, we corroborated the
suggestions based on MFM images that predicts the formation of three-domain
configurations along the hard axis, for temperatures above $25\,^{\circ}\mathrm{C}$.
The transition between this state and the configuration in which the
magnetization lies completely along the easy axis was also predicted,
but the value of the film aspect ratio for which this transition occurs
was far from the experimental one.

In addition, we compared the hysteresis loops that appears when an
external magnetic field is applied along the easy axis direction for
such three-domain configurations. Qualitatively, we obtained, using
an approach similar to Stoner and Wohlfarth \cite{stoner}, all the
three shapes of loops observed in the literature (squared, inclined
with large remanent magnetization and inclined with small remanent
magnetization). The main differences are that the inclined loops measured
experimentally are more rounded than ours and that the experimental
coercive fields are one order of magnitude smaller. 

Several factors help us to understand why all these differences between
the predictions of the model and the experimental measurements occur.
Firstly, the simple model we used here does not consider features
that are essential in the MnAs:GaAs real system, like the modulation
along the $y$ direction and the inter-stripe dipolar interaction.
To include these features, a more sophisticated model with bidimensional
domains and topological defects would be necessary. Moreover, in what
concerns the hysteresis loops, the Stoner-Wohlfarth method deals only
with the collective spin rotation, and does not take into account
nucleation or pinning, which can be responsible for the rounded shape
of the curves and the smaller coercive fields observed in the experiments.
As already pointed out in \cite{histerese_ploog}, due to the lack
of complete knowledge about the exact geometrical forms of the ferromagnetic
stripes, the materials inhomogeneities (that can induce nucleation)
and the correct microscopic parameters, we cannot expect an exact
reproduction of the experimental hysteresis loops. Nonetheless the
qualitative physical properties predicted here give an insight for
a more complete understanding of the complex domain structure of systems
like MnAs:GaAs films.

\section*{Acknowledgments}

The authors would like to thank CNPq and FAPESP for financial support.

\end{document}